# Regular Self-Pulsations in External Cavity Mid-IR Quantum Cascade Lasers

Nikola N. Vukovic, Jelena V. Radovanovic, Vitomir B. Milanovic, Alexander V. Antonov, Dmitry I. Kuritsyn, Vladimir V. Vaks, and Dmitri L. Boiko

*Abstract*—We show theoretically and experimentally that a quantum cascade laser (QCL) in an external Fabry-Pérot cavity (EC) can produce regular self-pulsations (SP) of the output intensity at frequencies ~75 GHz. We recognize that the propagation delay in EC provides QCL with a "memory" mechanism to preserve the regularity and coherence of the pulse train on the intervals significantly exceeding the sub-picosecond gain coherence and gain recovery times. These results may point to novel practical approaches for producing regular time-domain SPs and pulse trains in the Mid-IR QCLs. Otherwise, due to a multimode Risken–Nummedal–Graham–Haken (RNGH) instability, free-running FP QCLs exhibit SPs at Rabi flopping frequencies in the THz range, which in addition, reveal a quasi-periodic chaotic behavior if the cavity round trip time is a few picoseconds or more.

## I. INTRODUCTION

Periodic pulse production regimes such as self-Q-switching, self-pulsations or passive mode-locking (PML) in interband semiconductor laser diodes (LDs) are widely used in various practical applications. However these passive regimes are impossible in the mid-infrared (Mid-IR) Quantum Cascade Lasers (QCLs) because of their picosecond (or sub-picosecond) gain recovery time. This time is too short in comparison with the cavity round trip time and does not provide a "memory" mechanism (like the gain saturation on ns time intervals in LDs) to sustain such periodic regimes [1]. At the same time, the multimode Risken-Nummedal-Graham-Haken (RNGH) instability [2], [3] in QCLs with Fabry-Pérot (FP) cavities [1], [4] may pave a way to practical approaches for ultrafast pulse production in the Mid-IR range. More specifically, the spatial hole burning (SHB) effect in FP QCLs strongly reduces the excitation threshold for RNGH self-pulsations (SPs). However, there is no decent demonstration of regular (periodic) SPs in FP QCLs yet. (Section II provides a detailed review.)

In this paper, we report on using an external FP cavity (EC) as a delay line to obtain regular self-pulsations in QCL at a reduced repetition frequency. On one hand, the EC provides QCL with a round trip memory to sustain periodic pulse regimes. On the other hand it reduces the SP frequency itself. However, the EC naturally lowers the fractional cavity filling with the QCL gain medium and hence reduces the impact of the spatial hole burning on the lasing regime. Because the SHB is the key element required for occurrence of SPs in QCL, an over-lengthened EC would result in suppression of SPs.

We perform a numerical study on the effect of the EC length using a travelling wave rate equation model [5], [6]. In order to ease the numerical simulations we limit our study to a model system with a short QCL gain chip of 100 μm length having perfectly anti-reflective (AR) coated front facet. At first glance the considered EC QCL configuration may seem similar to the literature case of a laser operating under a *strong* optical feedback [7]. In fact the effect of optical feedback on QCL's dynamics is yet to be explored because very few theoretical and experimental studies are available in the literature [8]–[13] (see Section II). One recent communication [14] indicates a possibility to slightly tune the frequency of a self-generated QCL frequency comb by using a *weak* optical feedback. However these results do not agree with the report from Ref. [15] showing the single-mode operation of FP QCL under the optical feedback. In contrast to previous studies, here we consider the case of a *strong* optical feedback used to substantially modify the dynamics of SPs in multimode QCL.

Our numerical modelling predicts that EC QCL may reveal low-frequency regular SPs with the period (~13 ps) being substantially longer that the inherent QCL memory mechanism due to the gain coherence and gain recovery times (~0.1-1 ps). To validate our theoretical findings, we perform preliminary experimental tests of EC QCL and find a clear signature of regular intensity SPs at ~75 GHz frequency.

The paper is organized as follows. Section II provides a literature review on regular pulse train production in QCLs

This work was supported by Swiss National Science Foundation (SNF) project FastIQ, ref. no. IZ73Z0_152761, Ministry of Education, Science and Technological Development (Republic of Serbia), ref. no. III 45010, COST ACTIONs BM1205, MP1204 and MP1406, European Union's Horizon 2020 research and innovation programme (SUPERTWIN, ref. no. 686731), by the Canton of Neuchâtel and partially in the framework of state targets N 0035-2014-0206. *(Corresponding author: Nikola N. Vukovic.)*

N. N. Vukovic, J. V. Radovanovic, and V. B. Milanovic are with the School of Electrical Engineering, University of Belgrade, 11120 Belgrade, Serbia (e-mail: nikolavukovic89@gmail.com; radovanovic@etf.bg.ac.rs; milanovic@etf.bg.ac.rs).

A. A. Antonov, D. I. Kuritsyn, and V. V. Vaks are with the Institute for Physics of Microstructures RAS – Branch of FRC IAP RAS, GSP-105, 603950 N. Novgorod, Russia (e-mail: vax@ipmras.ru; aav@ipmras.ru; dk@ipm.sci-nnov.ru,)

D. L. Boiko is with the Centre Suisse d'Électronique et de Microtechnique (CSEM), CH-2002 Neuchâtel, Switzerland (e-mail: dmitri.boiko@csem.ch).

and enlightens why a search for self-sustaining (passive) regimes in QCLs is of vital importance. In Section III we describe our EC QCL model system. Section IV offers a discussion of the results of numerical simulations, while Section V presents preliminary experimental test and comparison with numerical simulations. Finally, Section VI provides a conclusion.

## II. OPERATING REGIMES OF QCLs

RNGH self-pulsations may occur in a continuous wave (CW) operating QCL due to a parametrically induced gain for non-lasing cavity modes [16]. The multimode instability is caused by Rabi oscillations that produce two parametric amplification sidebands being, respectively, red and blue shifted from the main lasing mode. The RNGH is manifested by a broadband multimode QCL emission with the overall spectral width being on the order of twice the Rabi flopping frequency. Due SHB induced carrier population grating, a few mm long FP QCLs exhibit multimode RNGH SPs at currents just a few percent above the lasing threshold [1], [4]. However the RNGH SPs in Mid-IR QCLs have not yet resulted in a reasonable experimental demonstration of a self-produced regular pulse train in the time domain [1], [4], [5], [17].

In the absence of group velocity dispersion in the cavity, the broad emission spectrum of RNGH SPs in QCL is composed of many regularly spaced cavity modes. So far such free-running [18] or stabilized [19], [20] QCL frequency combs have found numerous applications only in the spectral domain.

In our recent theoretical studies [5], [21], [22] we have demonstrated that regular RNGH self-pulsations can be obtained in stand-alone QCLs with sufficiently short FP cavities (~100 μm long). The regular periodic regime is favored because the cavity round trip time (~2 ps) is on the order of the gain recovery time [5]. Unfortunately, QCLs with such short cavities would require pulsed pumping at high current densities and exhibit impractically high pulse repetition rates in the THz range.

FP QCLs with a few mm long cavities operating in the regime of RNGH SPs are also of a limited use for the time–domain applications [1], [4], [17]. Their monolithic cavities are much longer than the coherence length [5]. When the pump current is turned on, different mutually incoherent domains along the cavity axis induce uncorrelated self-pulsations. Since the QCL gain medium does not provide a "memory" mechanism to sustain regular SPs on subsequent cavity round trips, the QCL reaches a steady-state RNGH regime of quasi-periodic chaotic SPs with multiple irregular oscillations on the cavity round trip [5] (see also the experimental evidence in [17]). As demonstrated in [23], the waveform of such chaotic pulsations depends strongly on the initial conditions. In this way neither short nor long single-section QCLs with monolithic FP cavities are suitable for practical realization of self-sustained ultrafast pulse production regimes in the Mid-IR range.

Important efforts to achieve pulse production in QCLs using an active mode-locking (AML) regime have also been made [24], [25]. However introduction of a monolithically integrated and separately contacted cavity section for active gain/loss modulation in QCL [24] has a limited use. Unlike a saturable electro-absorber in a mode-locked LD [26], the unipolar structure of QCL cannot be negatively biased to enhance the optical absorption. The QCL structure admits only the optical gain modulation via pump current in a single-section device [25] or via modulation of the pump current in a dedicated separately contacted gain section [24]. In fact, the most crucial element used in Ref. [24] for achieving periodic pulses in the time domain is the QCL structure itself. It operates on a diagonal transition with a long gain recovery time, although at the expense of the lasing efficiency, output power and cryogenic operation temperature. Here the gain saturation provides the QCL with the "memory" mechanism to sustain periodic oscillations on the subsequent round trips in the cavity. Yet the output pulses were not perfect in-terms of the peak to background ratio in the measured $2^{nd}$ order autocorrelation traces. We attribute the measured pulse imperfections to the tendency of mm-long FP QCLs to exhibit quasi-periodic chaotic SPs, favored by the SHB effect. An ideal approach for AML would be firstly to turn them into regular SPs or to totally suppress them.

According to the original studies of Risken and Nummedal, Graham and Haken [2], [3] the instability threshold in a ring unidirectional laser without SHB is significantly higher. Therefore placing a QCL gain chip in a ring cavity should provide a practical solution to avoid incoherent SPs and achieve a high-quality AML regime [25], [27],[23], [28]. As the counter propagating modes are just weakly coupled via backscattering, the mode competition in the gain medium forces a unidirectional lasing [29],[30]. However, in the experiment [25], the output pulses were surprisingly weak, of only 12 to 640 mW *peak* power. The pulse width was in the range from 75 ps to 4 ns, which is too large for the AML pulses considering the QCL gain bandwidth. In this respect we should recall that e.g. LD achieve their best ML performances under a hybrid mode-locking (HML) schemes [31] (and not in the AML regime), which implies that without active modulation, they operate in the passive mode-locking (PML) regime. *Following these considerations, we conclude that a search for alternative approaches to the passive regimes of periodic pulse production in Mid-IR QCLs is of vital importance, even for the applications with active modulation schemes*.

In the present paper we aim for regular SPs that we achieve by using an external FP cavity. The EC is used as a delay line to preserve the regular SP waveform.

## III. MODEL SYSTEM

As a model system for our study we use an external cavity quantum cascade laser schematically depicted in Fig. 1. We analyze its lasing dynamics as a function of the EC length and pump rate, assuming that the gain chip length is fixed at 100 μm. At first glance, its threshold current density is high, which may raise a question on the practical value of obtained results. However, a free-running 100 μm-long QCL is capable of producing regular self-pulsations. Therefore this model system has allowed us to study the effect of the EC length in a didactic manner and to identify a possible onset of chaotic dynamics caused specifically by the EC (as opposed to decoherence effects in a long gain chip). Using this model system, we have perceived how the EC slows down the rate of regular SPs and when EC may destabilize them.

In our model system, the front facet of the QCL chip facing the EC has an ideal anti-reflection (AR) coating (R=0%). The back facet of the chip is left uncoated ($R_1$= 27%). The cavity is closed by the output coupling mirror of the same reflectivity as for an "as-cleaved" facet ($R_2$= 27%). The propagation delay in the external cavity allows the laser to overcome the fast gain recovery time and provides our QCL system with a "memory" to sustain periodic regimes at lower repetition frequencies. The assumption about a complete suppression of reflections at the intra-cavity facet of the gain chip greatly simplifies numerical simulations. It is not crucial for validation of our concept as soon as the external cavity provides a strong optical feedback effect [7].

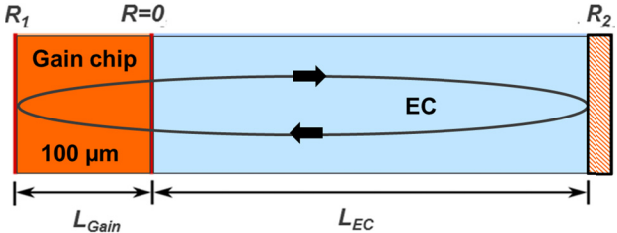

Fig. 1. Sketch of the EC QCL configuration used in our modeling. Note that the front chip facet is AR coated and that we account for the ultrafast coherent phenomena in the gain medium.

TABLE 1.
DYNAMIC MODEL PARAMETERS FOR INGAAS QCL CONSIDERED IN THIS PAPER.

| Symbol | Quantity | Value |
|---|---|---|
| $\lambda$ | Lasing wavelength | 10 μm |
| $T_1$ | Carrier lifetime | 1.3 ps |
| $T_2$ | Carrier dephasing time | 140 fs |
| $T_{2\_eff}$ | Effective carrier dephasing time in the presence of diffusion | 138 fs |
| $T_g$ | Relaxation time of the carrier population grating | 0.927 ps |
| $T_{2\_g}$ | Relaxation time of the coherence grating | 128 fs |
| $\alpha_i$ | Intrinsic material loss | 24 cm$^{-1}$ |
| D | Diffusion coefficient | 180 cm$^2$/s |
| $n_g$ | Group refractive index | 3.3 |
| $R_1$ | QCL back facet reflectivity | 27% |
| $R_{EC}$ | External reflector reflectivity | 27% |
| Γ | Optical mode confinement factor | 0.55 |
| ∂g/∂n | Differential material gain | 2.1×10$^{-4}$ cm$^3$/s |
| $n_t$ | Transparency carrier density | 7×10$^{14}$ cm$^{-3}$ |

The QCL gain chip parameters used in numerical simulations are listed in Table 1. We use a semiclassical travelling wave (TW) rate equation model [5] adapted for an external cavity laser. Following the approach from [6], we incorporate two cavity sections in our numerical TW model. One section provides a description for the QCL gain chip (carriers $N$ and coherences $P$) while another section describes an interaction-free propagation of an optical wave $E$ in the external cavity. Excitation of RNGH self-pulsation regime requires a perturbation [21], which is achieved by introducing polarization noise in the model (see model details in [5]).

## IV. MODELLING RESULTS

### A. Standalone monolithic FP QCL

As a reference system for our EC QCL study, we consider a standalone FP QCL chip with a very short cavity ($L_{cav}$=0.1 mm). The cavity round trip time ($T_{cav}$=2.2 ps) is comparable to the gain recovery time and QCL produces regular RNGH self-pulsations at a current of $p_{th2}$= 2.4 times above the lasing threshold [5], [21], [22]. Numerical simulations predict the onset of self-pulsations within the first 8 ps interval (~4 cavity round trips). After 30 ps (~14 cavity round trips), we observe a transit to a steady regime of regular self-pulsations [5]. (The waveform example can be found in the inset of Fig 2(a).) The field has a sine-wave envelope with a period being close to the cavity round trip time of 2.2 ps. The optical field and polarization of the gain medium change the sign at each passage in the cavity (at each half-period of the sine-wave). The intensity pulses are of 0.9 THz repetition rate and of 0.6 ps FWHM duration of 0.6 ps.

In contrast to our previous studies focused at the origin of the RNGH SPs [5], [21], [22], here we analyze bifurcations of the self-pulsations waveform as a function of the pump current, which we normalize to the lasing threshold $p=I/I_{th}$. We construct the bifurcation diagrams in a few different ways. In all cases the pump current is changed in small steps of $\delta p$=0.05. For each $p$ value, the numerical simulations are performed over 50 cavity round trips while only the last 25 periods after the end of all transients are used in the waveform analysis.

The bifurcation diagrams of extreme values (maxima and minima) in the output power waveform $P_{out}$ are plotted in Figs. 2(a) and (b). The red (blue) circles indicate the global and local maxima (minima) in the output power waveform. For instance, the largest values designate the peak power variations while the zero values in the output power waveform are due to the sign changes in the optical field envelop. (More details about such diagrams can be found in [32].)

In Fig 2(a), for each value of $p$, we reset the initial conditions to zero field and let the self-pulsations to occur and grow. Thus we obtain the bifurcation diagram of the self-starting SPs. For example, for $p$ above the second threshold $p_{th2}$=2.4, the QCL exhibits a self-starting oscillatory behavior, while below $p_{th2}$, it exhibits only CW emission. The insets in Fig. 2(a) display the corresponding output power waveforms for different pump rates and provide a hint for interpretation of the bifurcation diagram.

In the bifurcation diagram of Fig. 2(b), we study a stability of already ongoing SPs which have been excited earlier. In particular case, we start the numerical simulations from the largest value of $p$=5.5 to excite the SPs. After that we adiabatically reduce the pump current in small $\delta p$ steps. In contrast to the sequence used in Fig.2(a), for each value of $p$, we continue the simulations without resetting the initial conditions to zero field. Surprisingly, for the adiabatically decreasing pump rate, the SPs continue even below the self-excitation threshold $p_{th2}$. There is thus a hysteresis loop, associated with a bi-stable QCL operation either in the RNGH SPs or CW lasing regimes in the range from $p$= 1.5 to 2.4 [compare Figs.2(a) and 2(b)]. This hysteresis loop effect is in agreement with the results of Lyapunov stability analysis and RNGH threshold discussed in Refs [5], [21]: The SPs regime is possible, that is the Lyapunov exponent is positive, for $p>p_{min}$ ($p_{min}$≈1.3 was previously obtained for $L$=100 μm cavity case, see Fig. 2(a) in [5]). However the self-starting RNGH SPs in a short-cavity QCL can build up, provided the initial optical filed perturbation has a sufficient pulse area. This condition is met at $p>p_{th2}$ ($p_{th2}$≈2.4 in considered case). Note that the longer the QCL cavity, the smaller the hysteresis width. In mm-long QCL devices, the hysteresis loop practically disappears because $p_{th2}$≈$p_{min}$. [5], [21].

With increasing pump rate $p$, we observe a period doubling (at $p$=4) and a waveform distorsion of harmonic SPs. To study whether we can trigger a sequence of the period doubling and a bifurcation to a chaos, we resolve the QCL dynamics on the fast (within one cavity round trip) and the slow (on subsequent cavity round trips) time scales, while adiabatically changing $p$. In Fig. 2(c) we show a contour plot representing the evolution of the output power waveform with decreasing $p$. For 10>$p$>7.8, the period of SP in the output power waveform is equal to the half of the cavity round trip time. Then for 7.8>$p$>4, the period doubling occurs (the period of SP is equal to the round trip time). For $p$<4, the period of SP reduces back to the half of the cavity round trip. We thus do not observe any signature of period doubling sequence.

In Fig. 2(d) we calculate the normalized recurrence period density entropy (RPDE) [33], by applying a so-called time-delay embedding [34]. Recall that a perfectly periodic waveform (and a CW signal) has the RPDE entropy value of "0", while the RPDE value of "1" corresponds to a chaos. In this manner the RPDE indicates whether the SPs are periodic or quasi-periodic (chaotic).

For a standalone short-cavity QCL chip in Fig. 2(d), the RPDE entropy oscillates nearby "0" attesting for a perfectly periodic waveform even at high pump rate $p$, despite a strong anharmonicity. The spike on the RPDE curve at the lowest $p$ side of the SP/CW hysteresis loop can be attributed to the loss of stability of the SP regime, and, as a consequence, to a higher sensitivity of the waveform to polarization noise, which is continuously injected in the model system [5].

Hence, in the considered example of a standalone short-cavity QCL, the optical gain "memory" effect ($T_1$=1.3 ps) persists between the subsequent passage in the cavity ($T_{cav}$/2=1.1 ps), resulting in a regular pulse train. However, the pulse repetition frequency is very high.

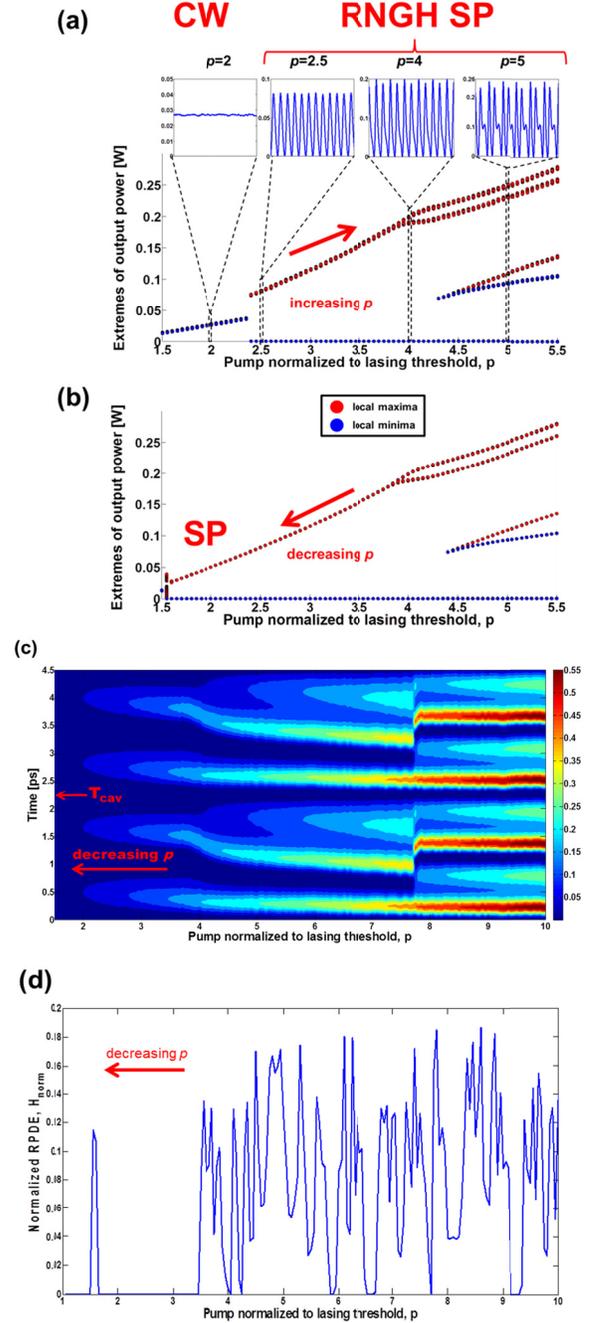

Fig. 2. Results of numerical simulations with TW model for monolithic QCL of the cavity length 100 μm: (a) Bifurcation diagrams of self-starting SPs represented by extreme values in the output power waveform $P_{out}$ (red and blue symbols correspond to the maxima and minima, respectively). The insets show the steady-state $P_{out}$ waveforms for different $p$ values. (b) Bifurcation diagram of stable SPs (but not necessarily self-starting) obtained by adiabatically decreasing $p$ from the initial value of $p$=5.5. Note a bi-stability region $p$=1.55 to 2.4 in the panels (a) and (b).  (c) Contour plot of $P_{out}$ distribution when decreasing $p$ (long time scale, $x$-axis), during time interval equal to 2 cavity round trips (short time scale, $y$-axis) (d) Behavior of the normalized Recurrence Period Density Entropy (RPDE) [33] with decreasing $p$. The following time-embedding parameters are used: embedding dimension $m$=4, embedding delay $\tau$=10, embedding ball radius r=0.003. More details about these parameters can be found in [33], [34].

## B. QCL in mm long EC

We now consider the same QCL gain chip ($L_G$=0.1 mm) placed in the 0.9 mm long external cavity with the overall cavity round trip time $T_{cav}$ of 8.2 ps. Note that without the SHB effect [21], the lasing threshold currents in our model systems with standalone QCL and EC QCL are equal.

In general, we observe that the (second) threshold current $p_{th2}$ for the excitation of SPs increases in the EC configuration. This is attributed to a reduced effect of SHB because of the fractional cavity filling with the QCL gain medium. The self-starting SPs occur at a higher second threshold of $p_{th2}\approx 4.3$. Figure 3 shows an example of the field amplitude, optical and RF power spectra as well as the 2$^{nd}$ order interferometric autocorrelation (IAC) trace at $p$=4.3. The dynamic behavior of EC QCL is very similar to the standalone QCL chip discussed in the previous section, including the sign change of the optical field at each half period of SPs (Fig. 3(a)) and negligible CW component (at zero Fourier frequency in Fig. 3(b)). However the frequency of SPs is reduced from 454 GHz down to 110 GHz close to the round trip frequency in extended cavity (the intensity oscillates at a twice higher frequency).

Like the standalone FP-cavity QCL, this EC QCL operates in the regime of RNGH SPs. Thus the main lasing mode at zero Fourier frequency in Fig. 3(b) is strongly suppressed, by more than 100 dB and the optical spectrum is dominated by the two adjacent sidemodes of the external cavity at +/-110 GHz (because of the mirror symmetry, the figure shows only the positive frequency part). The 2$^{nd}$ order IAC trace in Fig 3(c) has the peak-to-background ratio of 8:1.5 (as opposed to 8:1 for a perfectly periodic pulse train of low pulse duty cycle). We attribute the reduced IAC contrast to the pulse width being just a half of the SP period.

Like in the case of standalone QCL, we observe a SP/CW hysteresis loop between the domains of self-starting SP and CW operation as a function of the pump current $p$. Figure 4(a) shows the bifurcation diagram of extreme values of the output power obtained by decreasing $p$ starting from $p$=10 and Fig. 4(b) shows it as a contour plot of the output waveform resolved on the fast and slow time scales. In contrast to Fig 2, the fast time scale is set by the extended cavity round trip time $T_{cav}$. A comparison of the waveform evolution in the EC QCL (Fig 4) to standalone QCL chip (Fig.2) shows that RNGH SPs are more difficult to excite and they require higher pump rates $p$ (both $p_{min}$ and $p_{th2}$ are higher in EC QCL). This can be attributed to a partial cavity filling with the active gain medium and a reduced mode coupling via SHB effect.

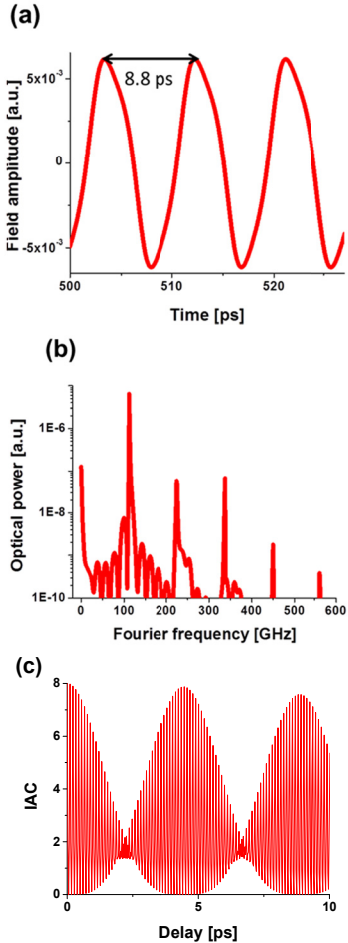

Fig. 3 EC QCL with the chip length of 100 µm and external reflector at 900 µm distance from the AR coated chip facet. Normalized pump rate is p=4.3. (a) Field amplitude waveform (b) Optical power spectra (positive frequency part, logarithmic scale) and (c) 2$^{nd}$ order interferometric autocorrelation function (IAC). Other QCL parameters are shown in Table 1.

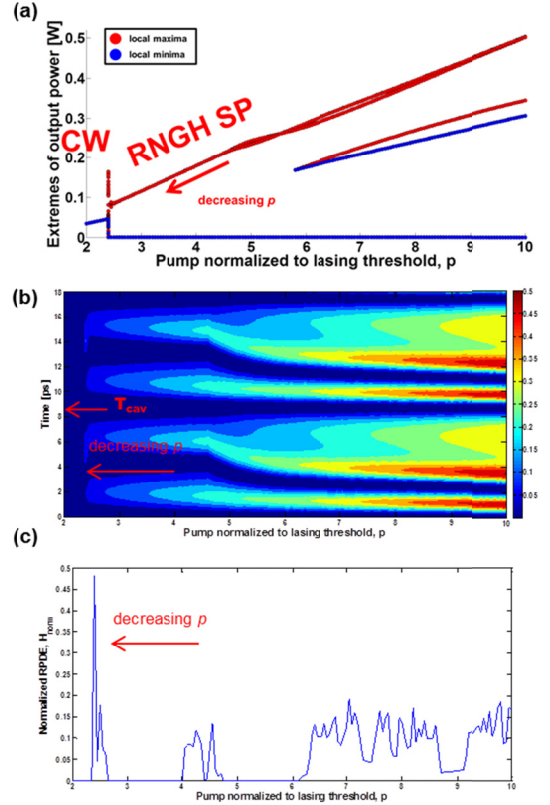

Fig. 4. EC QCL with the chip length of 100 µm and external reflector at 900 µm distance from the AR coated chip facet (same as in Fig. 3). (a) Bifurcation diagram of extremes of the output power waveform (red circles for $P_{max}$, blue for $P_{min}$) while decreasing normalized pump $p$ with 0.05 step. Panels (b) and (c) show the corresponding contour plot and normalized RPDE with the same embedding parameters as in Fig. 2(d) (m=4, τ=10, r=0.003).

At $p=p_{min}$, the SPs in both EC QCL ($p_{min}$ =2.5) and FP QCL ($p_{min}$=1.5) are at the cavity round trip time, and in both cases the period doubling occurs at higher $p$ (at $p>4.7$ for EC QCL and $p>4$ for FP QCL). Once again we attribute higher $p$ values needed for the period bifurcation to the smaller fractional cavity filling with the gain medium.

The evolution of the normalized recurrence period density entropy with the pump rate is displayed in Fig. 4(c). The low RPDE values attest to the regular periodic oscillations. The PRDE in EC QCL takes about the same values as in FP QCL [in Fig 2(d)], showing a few regions of perfect periodicity at reduced pump rates p. A solitary spike at $p=2.5$ is caused by the loss of stability of the SP regime, as discussed in relation to Fig 2(d).

We thus reach regular SPs despite the fact that their period is longer than the gain recovery time $T_1$ and that the active gain medium cannot provide a "memory" mechanism to sustain the periodic oscillations on multiple cavity round trips. These results can be interpreted in a way that the EC provides a "memory effect" to sustain periodic pulsations on the time intervals larger than $T_1$ although at the expense of a higher RNGH instability threshold. So far, in our model system we assume a 0.1 mm long chip. We may expect that for a longer chip (and higher cavity filling with the gain medium), the strength of mode coupling on SHB gratings will grow, lowering the SP threshold.

## C. QCL in cm-long EC

Being encouraged by the regular SPs at a reduced oscillation frequency in mm-long EC QCL, we examine if the EC length can be pushed further to the cm length scale. However, further increase of the external reflector distance to 9.9 mm from the perfectly AR coated chip facet (total round trip time $T_{cav}$ is of 68.3 ps) results in CW output in the whole range of the pumping rates representing any practical interest. Therefore, only this case from our numerical simulations is in agreement with the previous theoretical studies [11], [12] predicting enhanced stability of QCLs with an optical feedback.

Figure 5 shows the field amplitude waveform, optical spectra and 2$^{nd}$ order IAC trace when the pump rate is set a very high level of $p=27$. Under these conditions we do observe self-starting SP in the output power waveform. However these SPs are no longer of the RNGH type. The output power waveform has a strong CW component like in the case of a long unidirectional ring laser discussed in [35]. The main lasing mode dominates the optical spectrum. The absence of RNGH self-pulsations can be attributed to a too low cavity filling with the gain medium and respectively to a negligibly small mode coupling from the SHB effect. Although self-pulsations are periodic, the frequency of SPs is at the 5$^{th}$ harmonic of the external cavity frequency. The period of self-pulsations 13.5 ps does not match the propagation time neither in the gain chip nor in the EC. Such period fractioning contrasts with the period multiplication behavior of the RNGH SPs in FP QCL and mm-long EC QCL.

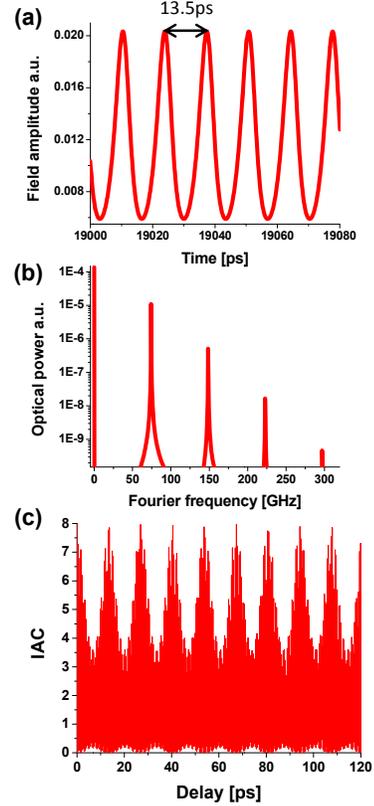

Fig. 5. EC QCL chip with the cavity length of 100 μm and external reflector placed at 9.9 mm distance from the AR coated chip facet. Normalized pump rate is p=27: (a) Field amplitude waveform (b) Optical power spectra (positive frequency part only, logarithmic scale) (c) IAC trace.

In Fig. 6 we plot the bifurcation diagrams of the extreme values of output power, waveform evolution contour plots and normalized RPDE with the changing pump rate.

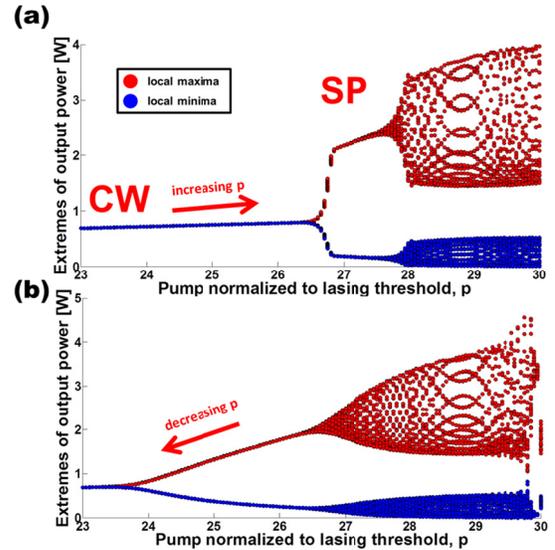

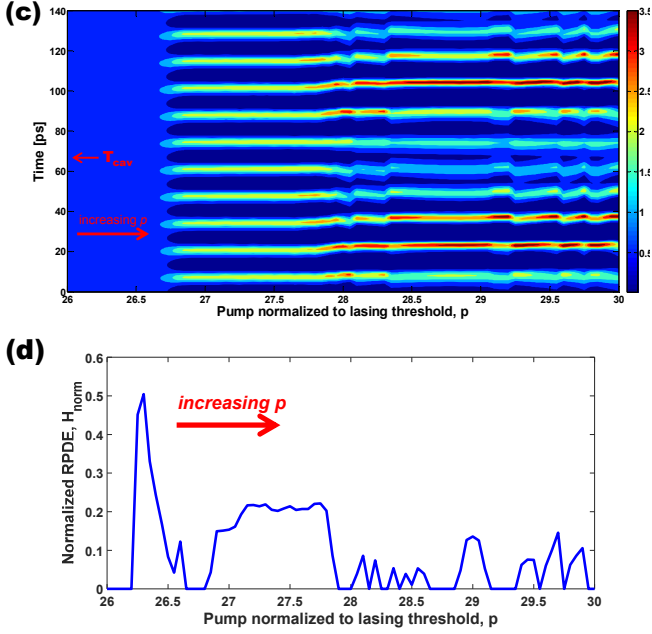

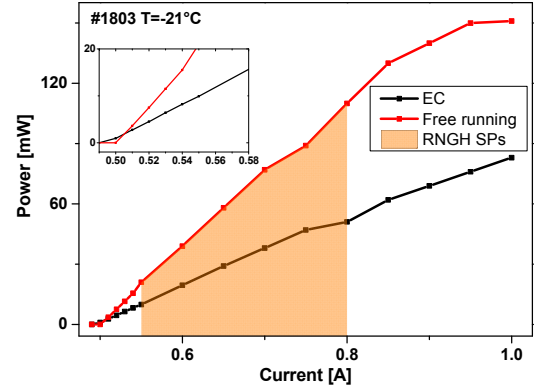

Fig. 6. EC QCL with the gain chip length of 100 μm and external reflector at 9.9 mm from the AR coated chip facet: (a) and (b) Bifurcation diagrams of extremes of output power waveform (red circles for $P_{max}$, blue for $P_{min}$) obtained with increasing and decreasing p, respectively. (c) Waveform evolution contour plot for increasing $p$ and corresponding (d) normalized RPDE obtained with the following embedding parameters: m=4, τ=10, r=0.003. Total number of periods per constant step is 10 and we cut the initial transient of the first 5 periods per step, without resetting field to zero at each new step (similar to Fig. 4).

With adiabatically increasing $p$, the self-starting SP occur at pump rates above $p$=26.5 [Fig. 6(a)], while adiabatic decrease of $p$ shows switching from SP to CW at a lower pump rate of $p$=23.5 [Fig. 6(b)]. Like in Fig.2 we therefore observe a bi-stable SP/CW behavior. However, for such AR coated QCL chip, the threshold of self-pulsations is prohibitively high for attempting any practical realization. We may thus conclude that SP are suppressed in cm-long EC QCL with 1% gain chip filling. Except for the transient at switching the SPs on, the RPDE entropy remains below 0.2 [Fig. 6(d)], indicating that the output power waveform has a regular character as opposed to a quasi-periodic (chaotic) signal. This is true even at very high pump rates $p$>28, when the waveform exhibits a multitude of the local maxima and minima as seen from Figs. 6(a) and (b). Figure 6(c) shows that in the range of $p$ between 26.5 and 27.7 SPs occur at a frequency close to $T_{cav}$/5. Nevertheless, for p>27.7 the period of SPs changes to a value close to one cavity round trip time $T_{cav}$. We conclude that in EC QCL with ~1 cm distance to the external reflector, RNGH SPs are impossible. The low frequency SPs may only occur at prohibitively high pump rates, but at a higher cavity filling with the gain chip, the excitation threshold is expected to lower.

## V. EXPERIMENTAL TEST AND DISCUSSION

In order to quickly grasp the implication of conclusions drawn from these numerical simulations, we perform a few preliminary experimental tests with QCL placed in external cavity. (A detailed experimental study with properly AR coated chips will be reported elsewhere.)

The QCL sample and the measurement techniques used in this study are detailed in [17], [36]. In brief, this is laterally single mode 3 mm long QCL chip emitting at ~8.2 μm (1226 cm$^{-1}$). The front facet has R=27% reflectivity and the back facet has $R_1$=95% coating (see Fig.1 for notation). The output is collimated with 4 mm focal distance ZnSe lens and the EC of 3 cm length is closed with $R_2$=50% reflector on a wedged ZnSe substrate with AR coated back side. The effective reflectivity of combined cavity mirror at resonance becomes $R_{eff} = \sqrt{R \cdot R_2} = 37\%$, which is expected to reduce the threshold current by 3%, i.e. by 15 mA for this device, which is a very minor change. Figure 7 shows the LI curves in the case of free running solitary QCL (red curve) and when it operates in EC configuration (black curve). The threshold current in EC configuration is indeed reduced by ~10 mA, attesting to a reasonable EC alignment (see the inset). Yet, the largest change that EC introduces is not in the threshold current $I_{th}$ but in the lasing regime.

Fig. 7. LI curves for chip #1803, where free running regime is denoted in red, while external cavity configuration is shown in black. Shaded region between 550 mA and 800 mA represents the region where RNGH self-pulsations occur (this is confirmed from both the optical spectra and 2$^{nd}$ order interferometric autocorrelation IAC).

In the range of currents 550-800 mA (the shaded region in Fig.7) the standalone QCL chip shows large spectral broadening and narrow SHG IAC correlation fringe with peak-to-background ratio of 8:3. Fig 8(a) shows an example of such SHG IAC trace. Note that the cavity round trip time of the free-running QCL chip is 66 ps (~15 GHz FSR) and the fringe after one cavity round trip exhibits a strongly reduced amplitude. This feature and the peak-to-background ratio of 8:3 evidence for an incoherent multimode regime with random relative phases of modes [37], [38]. Being accompanied by a broad optical emission spectrum (a comb) it should be attributed to quasi periodic (chaotic) RNGH self-pulsating behavior like the one observed in [1], [4], [17] and numerically modeled in [5] for mm-long FP-cavity QCLs. An example of a broad optical emission spectrum indicating the RNGH SPs is shown in Fig. 8(b). Its shape is remarkably different from the usual multimode behavior showing just a few competing modes.

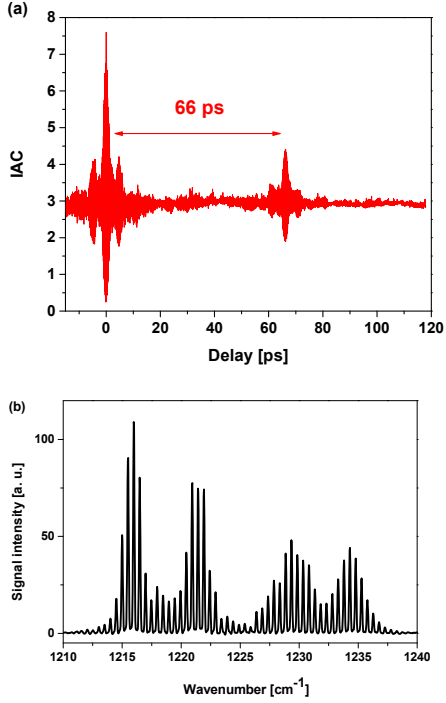

Fig. 8. Free-running QCL chip #1803 at 700 mA pump current and temperature -20°C: (a) 2$^{nd}$ order IAC trace showing narrow central fringe, secondary side-lobes, 8:3 peak to background contrast. (b) Broad emission spectrum or QCL frequency comb (linear scale). Both features attest to excitation of RNGH SPs.

However, when the same QCL gain chip was operated in the 3 cm long EC, we noticed a complete suppression of RNGH SPs in the entire range of pumping currents and excitation of just a few modes in the optical spectrum. This overall behavior is in reasonable agreement with the general trend seen in the EC QCL model system of Section IV.

For a few values of the pump current in the range of 600-750 mA, we observe low frequency self-pulsations. One example displayed in Fig. 9 shows the 2$^{nd}$ order IAC (a) and the optical spectra (b) measured in the EC QCL when the pump current is set to 600 mA. The overall round trip time in the extended cavity in this case is $T_{cav}$~266 ps (~3.8 GHz FSR). The noisy behavior in the measured 2$^{nd}$ order IAC can be attributed to the acoustic noise peak-up and instabilities of EC optical alignment during the delay time scanning but also to a lower output power (see LI curves in Fig.7)

The 2$^{nd}$ order IAC in Fig. 9(a) reveals periodic SPs. The pulse period is ~13 ps which corresponds to the pulse repetition frequency of ~77 GHz. This repetition frequency agrees reasonably with the separation of lasing modes in Fig. 9(b). The optical spectrum reveals the strong main lasing mode and the two optical side-modes shifted at ±2.5 cm$^{-1}$ frequencies (~75 GHz).

This low-frequency SPs regime is similar to the one predicted for the EC QCL model-system with a cm-long external cavity and depicted in Fig. 5. The low frequency of SPs and peculiar optical spectrum pattern evidence that this is no longer caused by the RNGH instability. In both cases in Figs. 5 and 9, the optical spectra are dominated by the main lasing mode and two symmetric sidemodes. The intensity SPs occurs due to beating of these modes, which thus have fixed phase relationships. The SP frequency is locked to a high order harmonic of the overall extended cavity round trip frequency because all lasing modes should be at corresponding cavity resonances. Along these lines we may conclude that, because of the partial reflections at the chip facet, SPs are observed only at certain currents when FP resonances of the gain chip and external cavity match each other. In both cases in Figs. 5 and 9, the external cm-long cavity lowers the SP frequency, enabling the pulse period to be significantly larger than the gain recovery time $T_1$. Interestingly, the 2$^{nd}$ order IAC fringe amplitudes in Fig. 9(a) after a delay of 66 ps (the round trip time in the chip) do not show a signature of strong phase degradation as in Fig. 8(a), implicating that the external cavity supplies a phase memory mechanism to sustain long –period SPs.

The fundamental difference between the cases in Figs. 5 and 9 is in the instability threshold level for these SPs. The longer QCL chip lowers the excitation threshold of this self-pulsations, due to a higher cavity filling by the gain medium and hence a stronger mode coupling via SHB gratings.

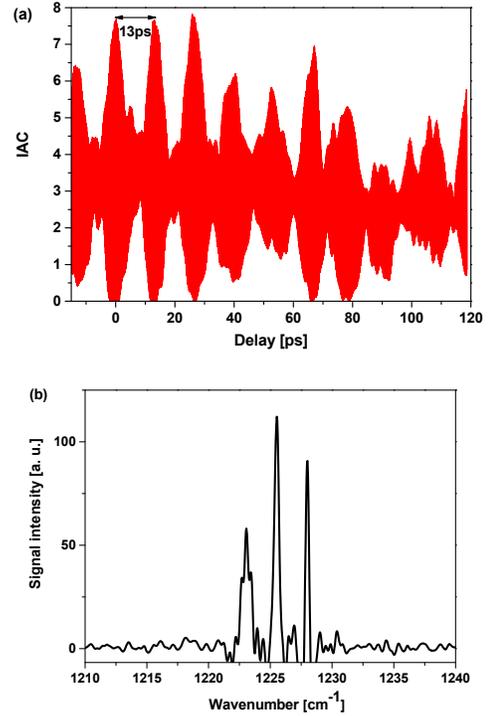

Fig. 9. QCL chip #1803 operating in EC of 3 cm long: (a) 2$^{nd}$ order IAC and (b) optical spectrum (linear scale) when the pump current is 600 mA and temperature -20°. Both feature attest to low frequency SPs. Compare to Fig.5.

Most of theoretical and experimental studies on optical feedback effect in QCL were focused, so far, at single-frequency distributed feedback (DFB) QCLs [8]–[11], [13], [39] and considered an adiabatic elimination of the medium polarization [12], thus excluding *a priori* the RNGH self-pulsations (and other ultrafast regimes) from the consideration.

Due to a negligible phase-amplitude coupling in QCL [10]–[12], single-frequency DFB QCLs subjected to a strong optical feedback reveal an enhanced stability of the CW lasing regime against the coherence collapse [15]. Mid-IR DFB QCLs exhibit coherence collapse just in a narrow range of the feedback strengths [8]. A recently published work [9] provides an evidence for the low-frequency (<<1 GHz) self-pulsations *at the EC frequency*. With further increased optical feedback strength, the DFB QCL bifurcates to low frequency quasi-periodic fluctuations and a chaos, followed by a steady CW operation [9], much like the text-book case of DFB LD subjected to the optical feedback [7]. None of these studies has considered the ultrafast coherent phenomena in the gain medium that may lead to multimode RNGH instability.

## VI. Conclusion

Owing to very short gain recovery times, free-running Mid-IR FP QCLs have gained a reputation of producing noisy multimode emission due to ultrafast quasiperiodic chaotic RNGH SPs at THz frequencies. For this reason passive regimes of regular SPs or ML are impossible in QCLs, while active modulation regimes have not yet resulted in demonstration of high-performance pulse production schemes.

In this paper we have analyzed behavior of QCLs in external cavities of different lengths, following the concept of using the propagation delay in EC to supply a "memory" mechanism and sustain a passive regime of regular self-pulsations. For cm-long EC QCL, we have demonstrated both theoretically and experimentally the excitation of such regular low-frequency (~75 GHz) self-pulsations at a high harmonic of the EC round trip frequency. The frequency of SPs and their regularity attest to the memory effect as a consequence of propagation in cm long EC, overcoming the negative effects caused by ultrashort gain recovery time in QCL.

Future work will focus on optimizing the cavity filling, chip facet coating and applying a current modulation to stimulate pulse production via a hybrid regime of SPs.

## Acknowledgment

The authors gratefully acknowledge the support of the funding agencies. QCL samples used in this research have been provided by Alpes Lasers.